\documentclass[journal]{IEEEtran}
\usepackage[noadjust]{cite}
\usepackage{siunitx}
\usepackage{amsmath}
\usepackage{graphicx}
\usepackage{xspace}
\usepackage[export]{overpic}
\usepackage{color}
\usepackage{xcolor}
\usepackage{soul} 
\soulregister\SI7
\soulregister\cite7
\soulregister\BiSe7
\soulregister\xspace7
\usepackage{comment}
\newcommand{\hle}[1]{\colorbox{yellow}{$\displaystyle#1$}}
\newcommand{\hldelete}[1]{\hl{#1}}

\newcommand{\BiSe}[0]{Bi$_2$Se$_3$\xspace}

\renewcommand{\hl}[1]{#1}
\renewcommand{\hle}[1]{#1}
\renewcommand{\hldelete}[1]{\iffalse{#1}\fi}
\renewcommand{\st}[1]{\iffalse{#1}\fi}

\ifCLASSINFOpdf
\else
\fi

\begin{document}
\title{Performance of Topological Insulator Interconnects}

\author{Timothy~M.~Philip,~\IEEEmembership{Student~Member,~IEEE,}
        Mark~R.~Hirsbrunner,
        Moon~Jip~Park,
        and~Matthew~J.~Gilbert,~\IEEEmembership{Member,~IEEE}% <-this % stops a space
\thanks{T. M. Philip, M. R. Hirsbrunner, and M. J. Gilbert are with the Department of Electrical and Computer Engineering, University of Illinois at Urbana-Champaign, Urbana, IL, 61802 USA (email: tphilip3@illinois.edu).}% <-this % stops a space
\thanks{M. R. Hirsbrunner and M. J. Park are with the Department of Physics, University of Illinois at Urbana-Champaign, Urbana, IL, 61802 USA.}% <-this % stops a space
\thanks{Manuscript received July 22, 2016; revised November 1, 2016.}}

\maketitle

\begin{abstract}
%The impending need for interconnect solutions for scaled integrated circuits beyond copper calls for the exploration of new materials.
The poor performance of copper interconnects at the nanometer scale calls for new material solutions for continued scaling of integrated circuits.
%Most recently, graphene nanoribbons (GNRs) have been proposed as a solution because of their linear energy dispersion, which results in high electron mobility.
%Using semiclassical methods, we show that finite width GNRs are sub-optimal for nanoscale interconnects as a result of their empirically observed bandgap reducing conductivity.
We propose the use of three dimensional time-reversal-invariant topological insulators (TIs), which host backscattering-protected surface states, for this purpose. Using semiclassical methods, we demonstrate that nanoscale TI interconnects have a resistance 1-3 orders of magnitude lower than copper interconnects and graphene nanoribbons at the nanometer scale. We use the nonequilibrium Green function (NEGF) formalism to measure the change in conductance of nanoscale TI and metal interconnects caused by the presence of impurity disorder. We show that metal interconnects suffer a resistance increase, relative to the clean limit, in excess of 500\% due to disorder while the TI's  surface states increase less than 35\% in the same regime. %The backscattering protection of TIs makes them much more resilient to the resistance increasing effects of scaling circuits than other state of the art materials, making them an excellent candidate for future nanoscale interconnects.
\end{abstract}


% \fi
\IEEEpeerreviewmaketitle



%\IEEEpubidadjcol

%=================================================================
%=================================================================
%=================================================================
\section{Introduction}

\IEEEPARstart{E}{lectronic} packaging is constantly evolving in order to achieve the lower power consumption and reduced circuit delays demanded by the scaling of microelectronic circuits. Copper is an effective solution for technology nodes in the near future, but finite-size effects in metals increase copper interconnect resistivity dramatically as dimensions decrease to \st{tens of nanometers}\hl{the nanoscale}~\cite{Steinhogl2005}.  By 2025, Metal 1 pitch is predicted to reach \st{10 nm }\hl{tens of nanometers}~\cite{Kapur2002}, where increased line edge roughness and grain boundary scattering in copper \hl{raise}\st{increase} resistivity, and thus dissipation and signal delay, to unacceptable values~\cite{Association2013,Graham2010}. A materials change will hence be necessary to avoid the ``interconnect bottleneck''~\cite{Ababei2005,Zhang2012}, whereby the poor performance  of nanoscale interconnects inhibits further scaling.

Zigzag graphene nanoribbons (GNRs) have been proposed as next generation \hl{interconnect} materials because of their high electron mobility~\cite{Bolotin2008}. Unfortunately, reliable fabrication of GNRs is difficult due to \st{random }defects in gro\hl{wth}\st{wn layers} and line edge roughness, \hl{both of }which\st{ both} increase scattering and degrade electron mobility~\cite{Hashimoto2004,Li2008}. In addition, finite-width GNRs develop a\st{ gap in their band structure}\hl{ band gap}~\cite{Son2006,Li2008,Zhang2010}, further reducing \st{its }\hl{their} conductance.

We propose the use of time-reversal-invariant topological insulators (TIs) for use in nanoscale interconnects. Topological insulators are a recently discovered class of materials that are gapped in their bulk spectrum but have surfaces that host massless, metallic Dirac fermions~\cite{Xia2009}. Time-reversal symmetry protects TI surface states from backscattering caused by charged disorder and edges, resulting in high conductance even in the presence of these scattering mechanisms. Therefore, TI interconnects will not suffer \hl{as much as copper interconnects and GNRs} from scaling-induced resistance increases\st{ as much as copper interconnects and GNRs}.

In this letter, we investigate the transport properties of metals, GNRs, and TIs to benchmark these materials for future nanoscale interconnects. We theoretically demonstrate that below \SI{6}{\nano\meter}, TI interconnect resistance is multiple orders of magnitude lower than the resistances of copper interconnects and GNRs due to the TI's backscattering protection. Using the non-equilibrium Green function (NEGF) formalism, we show that metal interconnects greatly increase in resistance with scaling-induced disorder, while transport through TI interconnects is comparatively insensitive. Having shown that TI interconnects continue to conduct well at the nanoscale while GNRs and copper do not, we conclude that TIs are excellent candidates for a future interconnect material.

%=================================================================
%=================================================================
%=================================================================

\section{Semiclassical Transport}

Although the widths of interconnects are decreasing, their lengths are often longer than the mean free path (MFP) of electrons and, as such, semiclassical calculations provide a useful picture of longitudinal transport~\cite{Niquet2014}. We use Matthiessen's rule to calculate the conductance of \st{copper, }the TI \BiSe\st{,} and GNRs, assuming that all scattering mechanisms are independent of each other~\cite{Bass1972}. We consider \BiSe because it is the most practical candidate for engineering purposes due to its bulk band gap of \SI{0.3}{\eV}~\cite{Xia2009}. Under Matthiessen's rule, the conductance is given by
\begin{equation}
  G = G_0 \sum_n \frac{1}{1+L\left(\Lambda^{-1} + \ell_n^{-1} \right)}, \label{eq:G}
\end{equation}
where $G_0$ is the conductance quantum, $L$ is the length of the wire, and $\Lambda$ is the experimentally measured, room temperature MFP. In this work, we use an MFP of \SI{1}{\micro\meter} for GNRs~\cite{Bolotin2008} and two MFPs of \SI{10}{\nano\meter} and \SI{100}{\nano\meter} for \BiSe~\cite{Chen2013,Wang2016}. The edge scattering length $\ell_n$ in \eqref{eq:G} is the distance that the $n$th transverse mode travels before scattering off an edge and is calculated from a modified form of the equation in~\cite{Naeemi2007}:
\begin{equation}
  \ell_n =
  % \begin{cases}
    % \displaystyle \frac{hv_F}{E_F E_g}\sqrt{E_F^2-E_g^2} & n = 0, \\
    \displaystyle W\sqrt{\frac{E_F^2 - E_g(W)^2}{\left(\hle{2\pi\hbar} v_Fn/2W\right)^2} -1}. \label{eq:ln}
  % \end{cases}
\end{equation}
Here, $W$ is the width of the wire, $E_F$ is the Fermi energy, $E_g(W)$ is the width-dependent band gap of the material, \hl{$\hbar$ is the reduced Planck constant,} and $v_F$ is \hl{the} Fermi velocity. The modification in \eqref{eq:ln} accounts for the band gap observed in both narrow GNRs and the surface states of thin \BiSe~\cite{Li2008,Zhang2010}. We use experimentally observed Fermi levels of \SI{0.26}{\eV} and \SI{0.21}{\eV} for \BiSe and  GNR, respectively~\cite{Xia2009,Berger2006}. Because the topological surface states of \BiSe are insensitive to scattering off crystalline edges~\cite{Alpichshev2010,Takane2012}, we exclude the $\ell_n$ term from its calculation. Additionally, the one-dimensional $n=0$ edge mode in GNRs is susceptible to weak localization making it nonconducting~\cite{Anderson1973}, thus we begin the sum in \eqref{eq:G} at $n=1$ for GNRs. \hl{We model the resistance of copper using a combined Fuchs-Sondheimer~\cite{Sondheimer1952} and Mayadas-Shatzkes~\cite{Mayadas1970} model for wires of aspect ratio 2~\cite{Association2013} to accurately capture both sidewall reflections and grain boundary scattering~\cite{Steinhogl2005}, resulting in values that agree well with experiments~\cite{Roberts2015}.  The integration of copper into CMOS manufacturing requires a diffusion barrier of a minimum width of 2 nm~\cite{Association2013}. In order to account for the liner in copper interconnects, we also calculate resistance where the line width includes the total added liner width of 4 nm. Since the diffusion coefficients for bismuth and selenium are orders of magnitude smaller than that of copper, it does not require a diffusion barrier for \BiSe~\cite{Hall1964,Kim1979,Ishikawa1989}.}

\begin{figure}
  \centering
  {\includegraphics[width=0.85\columnwidth]{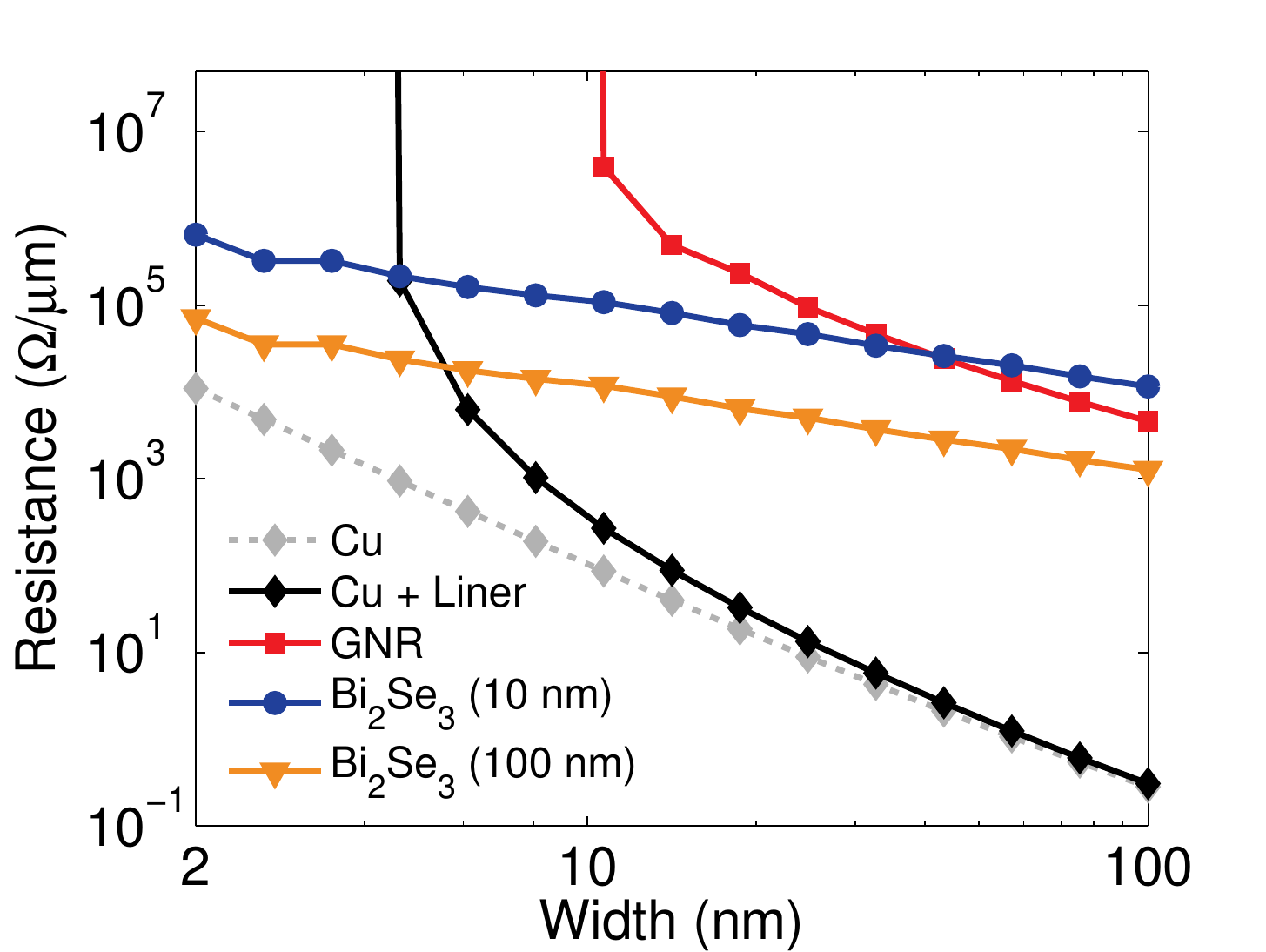}}
  % {\includegraphics[width=0.85\columnwidth,clip,trim={0 0.05in 0 0.32in}]{Plots/2016-10-10_MFP.pdf}}
  \caption{ The resistance of metallic GNRs, the TI \BiSe with MFPs of \SI{10}{\nano\meter} and \SI{100}{\nano\meter}, and copper \hl{with and without a diffusion liner}, each as a function of line width. \hl{Pure copper is the least resistive, but a required diffusion liner for manufacturing compatibility causes an resistance increase when included in line width.} For widths greater than \SI{6}{\nano\meter}, copper \hl{with liner} has a lower resistance than \BiSe and GNRs. Below a width of \SI{6}{\nano\meter}, the resistances of \BiSe are significantly lower than both copper \hl{with liner} and the GNR, regardless of the MFP. Copper's resistance rapidly grows due to the increased grain boundary \hl{and surface} scattering, while GNRs suffer from high resistance due to the increasing band gap in small-width ribbons.\label{fig:MFP}}
\end{figure}

Fig.~\ref{fig:MFP} illustrates that the \hl{lower resistance of pure copper makes it the optimal material for interconnects, but accounting for the required 2 nm diffusion liner shows it to be highly resistive below a physical interconnect width of} \SI{6}{\nano\meter}. \st{In this regime, }\hl{Above 6 nm}, \st{grain boundary}\hl{surface} scattering is insignificant, resulting in copper's high conductance. Despite having high mobilities, both GNRs and \BiSe are far more resistive than copper at this scale because their conductances are limited by their two-dimensional density of states. The especially poor performance of GNRs below \SI{40}{\nano\meter} is caused by the lack of an $n=0$ mode, which, combined with \st{its}\hl{their} band gap, significantly reduces the number of conduction channels\st{,} \hl{and} dramatically increases \st{its}\hl{their} resistance. Below \SI{6}{\nano\meter}, the resistance of copper increases rapidly, attributable to \st{the }increased \st{grain boundary density and associated}\hl{surface} scattering~\cite{Roberts2015}. We see that \BiSe with \st{both}\hl{either} MFP clearly outperforms copper and GNRs at this \hl{scale}\st{point} \hl{because it does not require a diffusion liner and has no edge scattering}.  

%%%%%%%%%%%%%%%%%%%%%%%%%%%%%%%%%%%%%%%%%%%%%%%%%%%%%%%%%%%%%%%%%%
%%%%    MFP (Figure 1)
%%%%%%%%%%%%%%%%%%%%%%%%%%%%%%%%%%%%%%%%%%%%%%%%%%%%%%%%%%%%%%%%%%
%%%%%%%%%%%%%%%%%%%%%%%%%%%%%%%%%%%%%%%%%%%%%%%%%%%%%%%%%%%%%%%%%%
%%%%%%%%%%%%%%%%%%%%%%%%%%%%%%%%%%%%%%%%%%%%%%%%%%%%%%%%%%%%%%%%%%

%=================================================================
%=================================================================
%=================================================================

\section{Influence of Disorder}

While Matthiessen's rule is useful for longer wires, we require a quantum description  for interconnect lengths below the MFP, where impurity-induced weak localization has a strong deleterious effect on conductance~\cite{Anderson19582}. We employ the NEGF formalism to understand the transport properties of materials below the MFP in the presence of disorder-induced, phase-coherent scattering. Using NEGF, we calculate the percent increase in resistance, relative to the clean limit, of both TI and metal interconnects as a function of impurity disorder strength. \hl{Although copper nanowires have been shown to have highly anisotropic conductance~\cite{Kharche2011,Hegde2014}, the resistance change due to uniform disorder is independent of direction~\cite{Li1996}, and thus we do not consider anisotropy here. We use TI and metal models that accurately display the qualitative transport characteristics of} \BiSe \hl{and copper but} \st{We }do not consider GNRs \st{here }because of their previously demonstrated insulating behavior at the nanoscale.

We use a Hamiltonian that accurately models the low energy behavior of TIs on a \st{three dimensional }cubic lattice, defined by \cite{Zhang2009,Liu2010,Hung2013}
\begin{equation}
\begin{split}
% \begin{split}
% 	H_\text{TI} &= \sum_{\mathbf{r}} \left[ \psi_\mathbf{r}^\dag \left(H_m+W_\mathbf{r}I_{4\times4}\right)\psi_\mathbf{r} \right. \\
% 	&\hspace{5em}+ \sum_{\boldsymbol\delta} \left. \left(\psi_\mathbf{r}^\dag H_{\boldsymbol\delta}\psi_{\mathbf{r} + \boldsymbol\delta} + \text{H.c.}\right) \right],
% \end{split} \\
	H_\text{TI} = &\sum_{\mathbf{r},\boldsymbol\delta} \psi_\mathbf{r}^\dag \left(H_m+\hle{d_\mathbf{r}}I_{4}\right)\psi_\mathbf{r} + \left(\psi_\mathbf{r}^\dag H_{\boldsymbol\delta}\psi_{\mathbf{r} + \boldsymbol\delta} + \text{H.c.}\right) , \\
  &H_m = M \Gamma^0, \qquad H_{\boldsymbol\delta} =  \frac{b\Gamma^0 + i\gamma\, \boldsymbol\delta \cdot \boldsymbol\Gamma}{2a^2}.
\end{split}\label{TI_H}
\end{equation}
The annihilation operator $\psi_{\mathbf{r}}$ is a \st{four component }spinor with two orbital and two spin degrees of freedom. The vectors $\boldsymbol\delta = (\pm a \,\mathbf{\hat x}, \pm a \,\mathbf{\hat y}, \pm a \,\mathbf{\hat z})$ are the distances between nearest neighbor atoms on the lattice, spaced by the lattice constant $a=\SI{3}{\angstrom}$. The matrices $\Gamma^{i}\, (i \in \{0, x, y, z\})$  are the Dirac gamma matrices, $\boldsymbol\Gamma =  (\Gamma^x\;\mathbf{\hat{x}},\Gamma^y\;\mathbf{\hat{y}},\Gamma^z\;\mathbf{\hat{z}})$, $I_4$ is the $4\times4$ identity matrix, and $M = m - 3b/a^2$. In this work, we set $m = \SI{1.5}{\eV}$, $b=\SI{9}{\eV\angstrom^2}$, and $\gamma = \SI{3}{\eV\angstrom}$ to put the insulator in the strong topological phase with a bulk band gap of \SI{1}{\eV}. \hl{This large band gap results in highly localized surface states that do not hybridize, allowing the simulation of smaller structures while maintaining the qualitative behavior of larger devices.} In \eqref{TI_H}, $\hle{d_\textbf{r}}$ is the disorder potential, which is randomly distributed in the range $\hle{-D/2 \leq d_\textbf{r} \leq D/2}$\hl{, representing impurities introduced during growth and fabrication~\cite{Zhu2016}. Conductance calculations are averaged over ten trials for each disorder strength $D$, where each trial has a different random disorder potential configuration.} The disorder range studied, $\SI{0}{\eV}  \leq \hle{D}  \leq \SI{5}{\eV}$, corresponds to a surface state MFP down to \SI{0.32}{\nano\meter}, using the relation $\Lambda = 12 \hbar^3 v_F^3/(a^2 \hle{D^2E_F} )$~\cite{Akera1990}\hl{, covering the range of experimentally measured MFPs in TIs~\cite{Chen2013,Wang2016}.} The metal is modeled by a \st{simple }3D tight-binding Hamiltonian with nearest-neighbor hopping $t_0 = \SI{1.5}{\eV}$~\cite{Datta2000}, such that the metal has the same bandwidth as the TI. The chemical potential is set to \SI{.8}{\eV}, \hl{although qualitative trends are independent of specific value}. Random \st{on-site }impurity disorder is added to the metal in the same fashion as for the TI. \hl{Grain boundary scattering is not relevant here as the device dimensions are smaller than typical grain sizes.} Both materials are modeled using a wire with dimensions $\left(10a \,\mathbf{\hat{x}}, 5a \,\mathbf{\hat{y}}, 5a \,\mathbf{\hat{z}}\right)$,\st{ with} where transport \hl{is} simulated along $\mathbf{\hat{x}}$\hl{ with a bias of 1 mV and }\st{. The}temperature\st{ is set to} at \SI{300}{\K}\st{, and a bias of 1 mV is applied along the transport direction}.

%  \hl{The surface states of the TI model do not gap out at these dimensions because the surface states are much more localized than in \BiSe.} 
%\begin{equation}
    % H_\text{metal} = \sum_{\mathbf{r}} \left[\left(6t_0 + W_\mathbf{r}\right)\psi_\mathbf{r}^\dag \psi_\mathbf{r} - t_0 \sum_{\boldsymbol{\delta}} \left(\psi_\mathbf{r}^\dag \psi_{\mathbf{r} + \boldsymbol\delta} +\text{H.c.}\right) \right],  
%   H_\text{metal} = \sum_{\mathbf{r},\boldsymbol{\delta}} \left[\psi_\mathbf{r}^\dag\left(6t_0 + W_\mathbf{r}\right) \psi_\mathbf{r} -\left(\psi_\mathbf{r}^\dag  t_0\psi_{\mathbf{r} + \boldsymbol\delta} +\text{H.c.}\right) \right],  
%\end{equation}
%%%%%%%%%%%%%%%%%%%%%%%%%%%%%%%%%%%%%%%%%%%%%%%%%%%%%%%%%%%%%%%%%%
%%%%    Resistance Increase (Figure 2)
%%%%%%%%%%%%%%%%%%%%%%%%%%%%%%%%%%%%%%%%%%%%%%%%%%%%%%%%%%%%%%%%%%
\begin{figure}
  \centering
  % \begin{overpic}[width=0.83\columnwidth,clip,trim={0 0.00in 0 0.15in}]{Plots/DisorderDeltaR.pdf}
  % \put(18.5,35){{\includegraphics[scale=.40,clip,trim={4.6in 2.85in 6.6in 2.1in}]%
  % {Plots/TI_band_cartoon.pdf}} } 
  % \end{overpic}
  \begin{overpic}[width=0.75\columnwidth,clip,trim={0 0 0 0}]{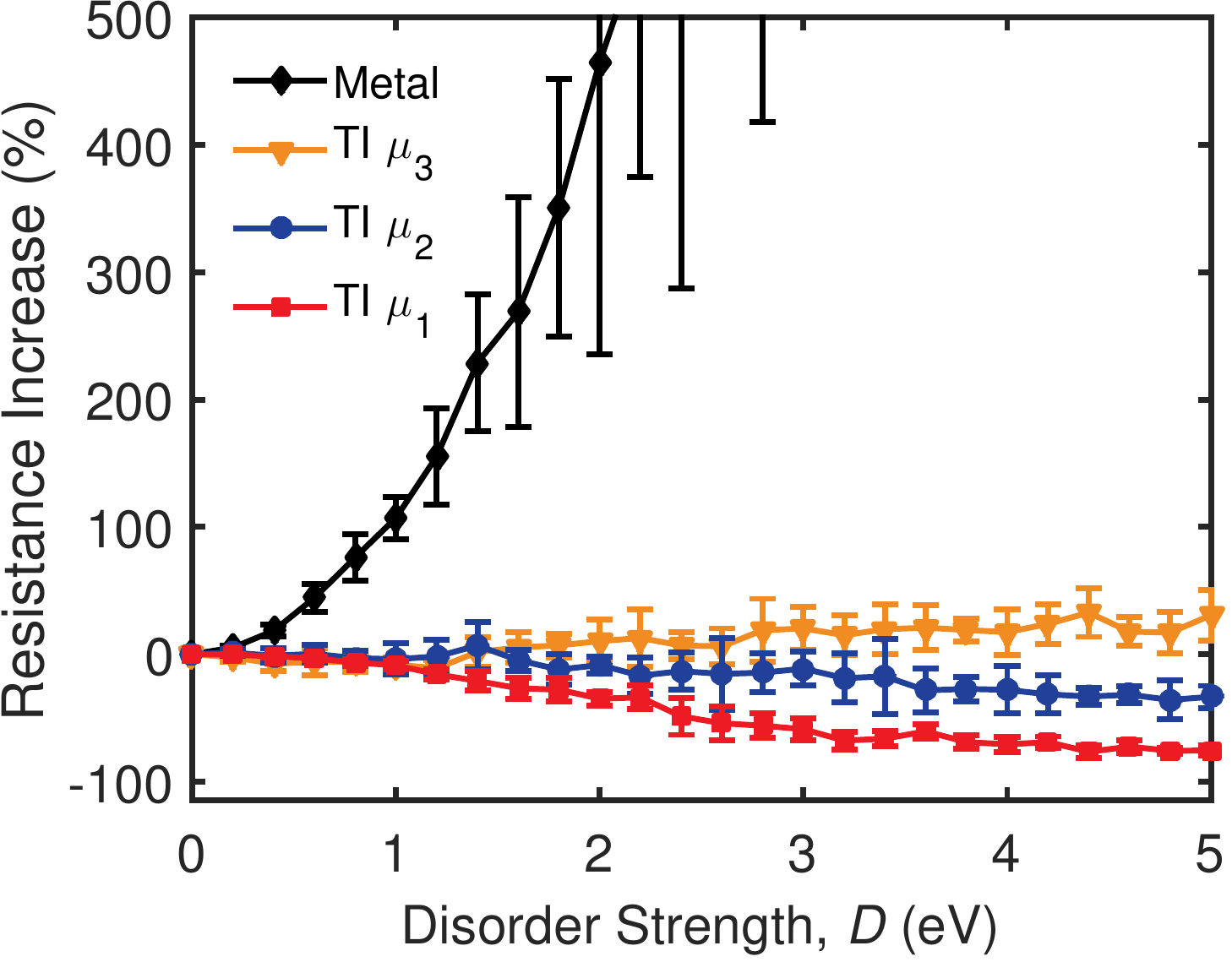}
  \put(65,36){{\includegraphics[scale=.40,clip,trim={4.6in 2.85in 6.6in 2.1in}]%
  {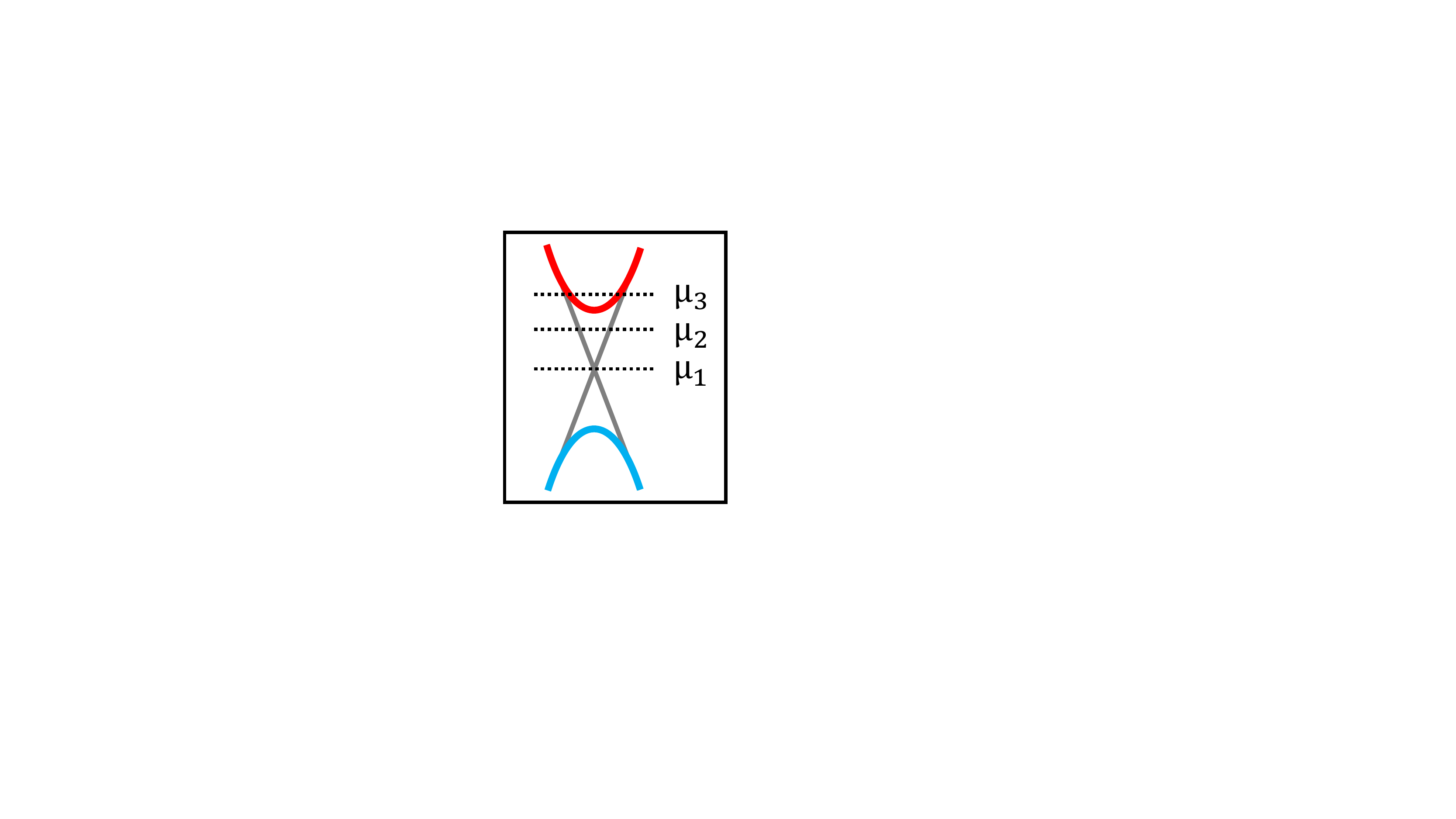}} } 
  \end{overpic}
  % \caption{Percent increase in resistance, relative to the clean limit, of a metal and a TI with chemical potentials [$\mu_1 = 0$, $\mu_2 = \SI{0.85}{\eV}$, and $\mu_3 = \SI{1.50}{\eV}$] as a function of disorder, simulated using NEGF. Conductance is \hl{averaged over ten trials, each using a different random disorder potential configuration}, resulting in \hl{error bars}\st{standard deviations} too small to display\st{ using error bars}. The inset graphic depicts the TI's band structure and the different chemical potentials used, denoted with dotted horizontal lines. Because of the symmetry protection of the TI's surface states, there is little change in the TI's resistance, relative to the metal's. The metal's resistance increases immediately, quickly becoming an insulator. \label{fig:disorder}}
  \caption{Percent increase in resistance, relative to the clean limit, of a metal and a TI with chemical potentials $\mu_1 = 0$, $\mu_2 = \SI{0.3}{\eV}$, and $\mu_3 = \SI{0.6}{\eV}$ as a function of disorder, simulated using NEGF. Conductance is \hl{averaged over ten trials, each using a different random disorder potential configuration}\st{, resulting in standard deviations too small to display using error bars}. The inset graphic depicts the TI's band structure and the different chemical potentials used, denoted with dotted horizontal lines. Because of the symmetry protection of the TI's surface states, there is little change in the TI's resistance, relative to the metal's. The metal's resistance increases immediately, quickly becoming an insulator. \label{fig:disorder}}
\end{figure}

\begin{figure}
	\centering
  % {\includegraphics[width=0.83\columnwidth,clip,trim={0 0.02in 0 0.36in}]{Plots/TI_transmission.pdf}}
  {\includegraphics[width=0.72  \columnwidth,clip,trim={0 0 0 0}]{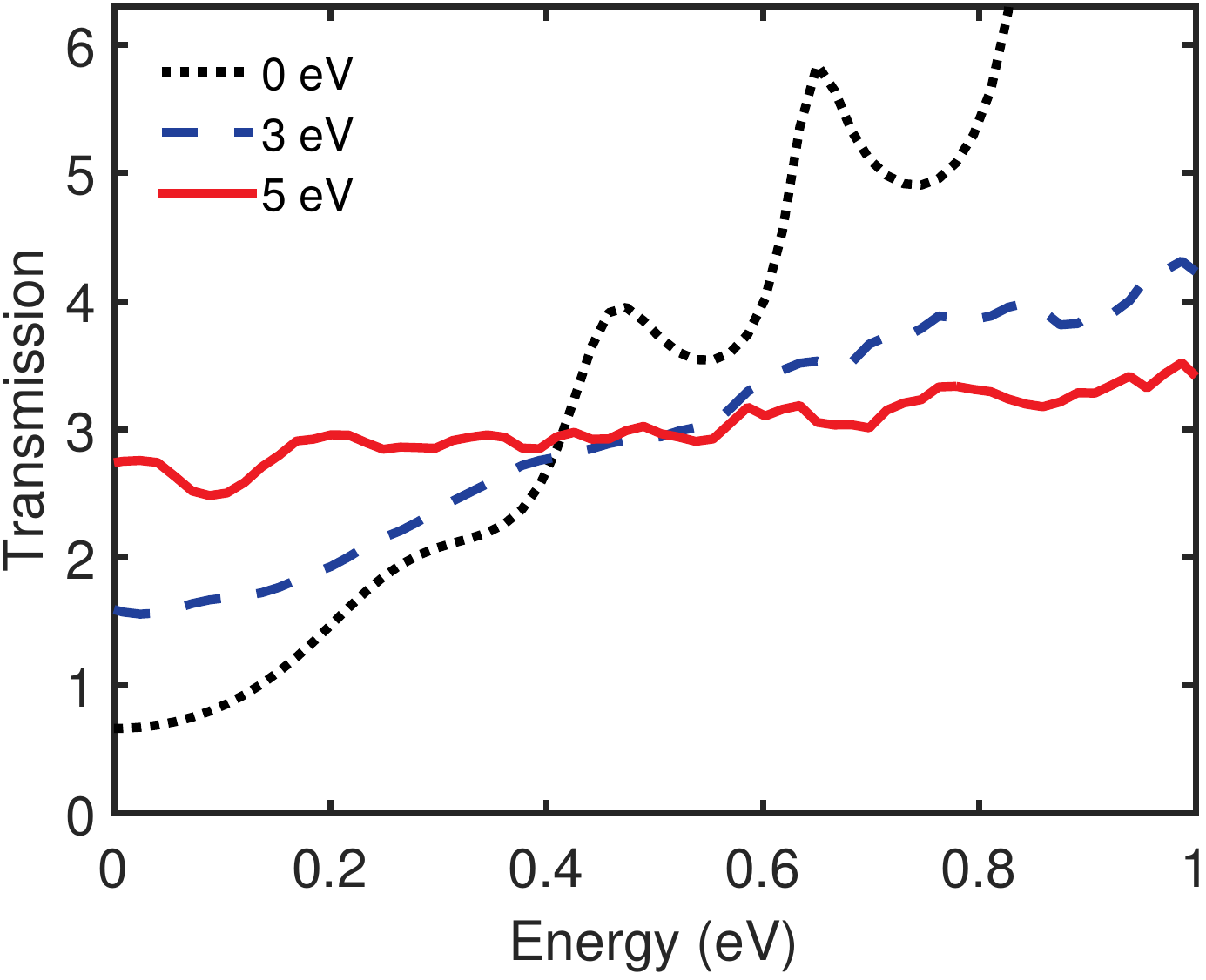}}
  % \caption{ The average transmission function of the TI interconnect for $D = \SI{0}{\eV}$, $\SI{1.1}{\eV}$, $\SI{2.2}{\eV}$, and $\SI{3.3}{\eV}$. Transmission is symmetric about $E = \SI{0}{\eV}$ as a result of particle-hole symmetry in the model. As the TI transitions to a diffusive metal phase, conductive mid-gap states form that improving low energy transmission. At higher energies, however, disorder-induced weak localization continuously reduces transmission with increasing disorder, thus increasing resistance of bulk transport.\label{TI_trans}}
  \caption{ The disorder-averaged transmission function of the TI interconnect for $D = \SI{0}{\eV}$, $\SI{3}{\eV}$, and $\SI{5}{\eV}$. Transmission is symmetric about $E = \SI{0}{\eV}$ as a result of particle-hole symmetry in the model. As the TI transitions to a diffusive metal phase, conductive mid-gap states form, improving low energy transmission. At higher energies, however, disorder-induced weak localization continuously reduces transmission with increasing disorder, thus increasing resistance of bulk transport.\label{TI_trans}}
\end{figure}

Fig.~\ref{fig:disorder} shows the percent increase in resistance, relative to the clean limit, of each interconnect versus disorder strength\st{ $D$}. For the metal interconnect, on-site impurity disorder increases elastic scattering, resulting in more than a \hl{450\%} increase in resistance above \hl{2~eV} of disorder. \st{Furthermore, }We plot the resistance of the TI at three different chemical potentials ($\mu_1$, $\mu_2$ and $\mu_3$), illustrated by the inset of Fig.~\ref{fig:disorder}. For conduction through the Dirac point at $\mu_1$, \st{the resistance decreases slightly in the presence of disorder. This decrease is caused by}\hl{the presence of disorder decreases resistance by 76\% at $D= 5$ eV.} Disorder-induced mid-gap states \st{increasing}\hl{increase} the number of conduction channels\hl{, as is evident in Figure~{\ref{TI_trans}}, resulting in } the TI transitioning into a diffusive metal phase~\cite{Kobayashi2013a,Sbierski2014,Kobayashi2014}. Transport at \st{chemical potential }$\mu_2$, crossing solely through the TI surface states, results in \st{only a 150\% increase}\hl{a slight decrease} in resistance \st{because of the backscattering protection of the surface channels}\hl{as transport occurs at higher energies than most of the disorder-induced mid-gap states}. \st{No metallic transition occurs because $\mu_2$ is well above the energy level of disorder-induced mid-gap states.}\hl{The small change in resistance for the TI at $\mu_1$ and $\mu_2$ compared to the dramatic rise for the metal demonstrates the benefit of the topological protection of the TI surface states. } \st{At c}Chemical potential $\mu_3$\st{, which} crosses both the surface states and the bulk bands, \hl{which results in} the resistance \st{steadily increases with disorder strength}\hl{increasing by 30\% at $D=5$ eV due to the localization of the unprotected bulk electrons. Continued conduction through the surface states, however, limits the resistance increase in the TI.}\st{. However, above 4.5~eV, the resistance saturates at a 450\% increase, as only the topologically protected surface states remain conducting.} Because the Fermi level of as-grown \BiSe crosses both the bulk and surface bands~\cite{Xia2009}, our calculations at $\mu_3$ are of particular interest as they indicate that \BiSe can benefit from both bulk conductance and surface backscattering protection.

\section{Additional Considerations}
Recent theoretical work suggests that inelastic scattering by acoustic phonons greatly reduces the mobility of TIs~\cite{Gupta2014}. Although this is true for long lengths, TI interconnects with lengths less than or equal to the MFP may not suffer such a large degradation. To investigate inelastic scattering, we add a phenomenological on-site \st{imaginary }self-energy to the TI Hamiltonian. The scattering self-energy $\Sigma_S = -i\hbar/2\tau$~\cite{Datta2000} is characterized by the mean free time $\tau  = \Lambda/v_F $. Here, the MFP $\Lambda$ is \SI{23.8}{\nano\meter}~\cite{Wang2016} and the Fermi velocity $v_F$ is \SI{5e5}{\meter/\second}~\cite{Zhang2009},  resulting in $\tau = \SI{47.6}{\femto\second}$ and a scattering self-energy  $\text{Im}\{\Sigma_S\} = \SI{-.014}{\eV}$. \st{Transport simulated using NEGF}\hl{Simulated transport} shows only a 20\% resistance increase over \st{transport}\hl{that} without inelastic scattering, a much smaller reduction than was reported in~\cite{Gupta2014}. As such, we see that inelastic phonon scattering is not a significant source of performance degradation in nanoscale TI interconnects.

Another concern in the use of TI interconnects is that they would be in the presence of time-reversal-breaking electromagnetic fields from nearby lines, which could destroy their topological protection. We estimate the influence of this crosstalk by using Ampere's law for a wire carrying a current of \SI{1}{\milli\ampere} at an interconnect pitch of \SI{5}{\nano\meter} resulting in a magnetic field strength $|\mathbf{B}| = \SI{40}{\milli\tesla}$. Such a field creates a Zeeman energy gap  $E_Z = g \mu_B |\mathbf B|$, where $g \approx 32$ is the g-factor for \BiSe~\cite{LandoltBornstein1998} and $\mu_B = $ \SI{57.9}{\micro\eV\ensuremath{\cdot}\tesla} is the Bohr magneton. Using this relation, we obtain a Zeeman energy splitting of \SI{74.1}{\micro\eV}. Therefore, even in the presence of many other \hl{lines}\st{ interconnects}, this gap will be smaller than \SI{1}{\milli\eV}, resulting in an immeasurable impact on the TI's topological properties.

%%%%%%%%%%%%%%%%%%%%%%%%%%%%%%%%%%%%%%%%%%%%%%%%%%%%%%%%%%%%%%%%%%
%%%%    TI Transmission (Figure 3)
%%%%%%%%%%%%%%%%%%%%%%%%%%%%%%%%%%%%%%%%%%%%%%%%%%%%%%%%%%%%%%%%%%
%%%%%%%%%%%%%%%%%%%%%%%%%%%%%%%%%%%%%%%%%%%%%%%%%%%%%%%%%%%%%%%%%%
%%%%%%%%%%%%%%%%%%%%%%%%%%%%%%%%%%%%%%%%%%%%%%%%%%%%%%%%%%%%%%%%%%

%=================================================================
%=================================================================
%=================================================================
\section{Conclusion}

We have performed a numerical study to explore the use of TIs as future interconnects. Using semiclassical techniques, we find that copper is much less resistive than the Bi$_2$Se$_3$ or GNRs above line widths of \SI{6}{\nano\meter}. Below this width, however, the increased \st{grain boundary}\hl{surface} scattering in copper and the observed band gap in GNRs cause both to rapidly rise in resistance above Bi$_2$Se$_3$, making the TI the best candidate in this regime. Using NEGF, we also observe that disorder causes the metal's resistance to increase by orders of magnitude\hl{ but has no negative impact on }the TI's backscattering-protected surface states\st{ only increase two-fold}. Because TIs maintain their conductive properties under the effects of scaling microelectronics, they are excellent candidates for next-generation interconnect materials.

\section*{Acknowledgment}

T. M. Philip and M. J. Park would like to thank Y. Kim for helpful discussions. The authors would like to thank NSF CAREER ECCS-1351871 for funding.

\ifCLASSOPTIONcaptionsoff
  \newpage
\fi



%\bibliographystyle{IEEEtran}


% \end{thebibliography}
\bibliographystyle{IEEEtran}
% \bibliography{../../../../Latex_Headers/bibtex/Interconnects}
\bibliography{IEEE_EDL_TI_Interconnect_v2_1}

\end{document}